\pdfoutput=1
\RequirePackage{ifpdf}
\ifpdf % We are running pdfTeX in pdf mode
\documentclass[pdftex]{sigma}
\else
\documentclass{sigma}
\fi

\numberwithin{equation}{section}

\newtheorem{Theorem}{Theorem}[section]
 { \theoremstyle{definition}
\newtheorem{Definition}[Theorem]{Definition}
\newtheorem{Example}[Theorem]{Example} }

\newcommand{\calL}{{\cal L}}
\newcommand{\calM}{{\cal M}}
\newcommand{\md}{{\rm d}}
\newcommand{\bracket}[2]{\langle #1,#2\rangle}

\begin{document}
\allowdisplaybreaks

\newcommand{\arXivNumber}{1905.02434}

\renewcommand{\PaperNumber}{076}

\FirstPageHeading

\ShortArticleName{Momentum Sections in Hamiltonian Mechanics and Sigma Models}

\ArticleName{Momentum Sections in Hamiltonian Mechanics\\ and Sigma Models}

\Author{Noriaki IKEDA}

\AuthorNameForHeading{N.~Ikeda}

\Address{Department of Mathematical Sciences, Ritsumeikan University, Kusatsu, Shiga 525-8577, Japan}
\Email{\href{mailto:nikeda@se.ritsumei.ac.jp}{nikeda@se.ritsumei.ac.jp}}

\ArticleDates{Received May 24, 2019, in final form September 29, 2019; Published online October 03, 2019}

\Abstract{We show a constrained Hamiltonian system and a gauged sigma model have a~structure of a momentum section and a Hamiltonian Lie algebroid theory recently introduced by Blohmann and Weinstein. We propose a generalization of a momentum section on a pre-multisymplectic manifold by considering gauged sigma models on higher-dimensional manifolds.}


\Classification{53D20; 70H33; 70S05}

\section{Introduction}

Recently, relations of physical systems with a Lie algebroid structure and its generalizations have been found and analyzed in many contexts. For instance, a Lie algebroid~\cite{Mackenzie} and a~generalization such as a Courant algebroid appear in topological sigma models \cite{Ikeda:2012pv}, T-duality~\cite{Cavalcanti:2011wu}, quantizations, etc.

Blohmann and Weinstein \cite{Blohmann:2018} have proposed a generalization of a momentum map and a~Hamiltonian $G$-space on a Lie algebra (a Lie group) to Lie algebroid setting, based on analysis of the general relativity \cite{Blohmann:2010jd}. It is called a momentum section and a Hamiltonian Lie algebroid. This structure is also regarded as reinterpretation of compatibility conditions of geometric quantities such as a metric $g$ and a closed differential form $H$ with a Lie algebroid structure, which was analyzed by Kotov and Strobl~\cite{Kotov:2016lpx}.

In this paper, we reinterpret geometric structures of physical theories as a momentum section theory, and discuss how momentum sections naturally appear in physical theories. Moreover, from this analysis, we will find a proper definition of a momentum section on a pre-multisymplectic manifold.

We analyze a constrained Hamiltonian mechanics system with a Lie algebroid structure discussed in the paper~\cite{Ikeda:2018rwe}, and a~two-dimensional gauged sigma models \cite{Hull:1990ms} with a $2$-form B-field and one-dimensional boundary. In a constrained Hamiltonian mechanics system, we consider a~Hamiltonian and constraint functions inhomogeneous with respect to the order of momenta. Then, a zero-th order term in constraints is essentially a momentum section. In a~two-dimensional gauged sigma model, a pre-symplectic form is a B-field, and a~one-dimensional boundary term is a momentum section. Two examples are very natural physical systems, thus, we can conclude that a momentum section is an important geometric structure in physical theories.

Recently, a two-dimensional gauged sigma model with a $2$-form B-field with three-dimensional Wess--Zumino term \cite{Hull:1990ms} is analyzed related to T-duality in string theory \cite{Buscher:1987sk, Buscher:1987qj, Chatzistavrakidis:2015lga, Chatzistavrakidis:2016jci, Chatzistavrakidis:2017tpk, Chatzistavrakidis:2016jfz}. For such an application, it is interesting to generalize a momentum section in a pre-multisymplectic manifold. The string theory sigma model with an NS $3$-form flux $H$ is the pre-$2$-symplectic case in our theory.

In this paper, we consider an $n$-dimensional gauged sigma model with $(n+1)$-dimensional Wess--Zumino term. The Wess--Zumino term is constructed from a closed $(n+1)$-form $H$, which defines a pre-$n$-plectic structure on a target manifold~$M$. For gauging, we introduce a vector bundle $E$ over $M$, a connection $A$ on a world volume $\Sigma$ and a Lie algebroid connection~$\Gamma$ on a~target vector bundle~$E$. Consistency conditions of gauging give geometric conditions on a series of extra geometric quantities $\eta^{(k)} \in \Omega^k\big(M, \wedge^{n-k}E^*\big)$, $k= 0, \dots, n-1$. From this analysis, we propose a definition of a momentum section on a pre-multisymplectic manifold. This definition comes from a natural physical example, a gauged sigma model. We see that our definition of a~momentum section on a~pre-multisymplectic manifold is a generalization of a momentum map on a multisymplectic manifold \cite{Carinena:1992rb, Gotay:1997eg}.

This paper is organized as follows. In Section~\ref{section2}, we explain definitions of a momentum section and a Hamiltonian Lie algebroid. In Section~\ref{section3}, we show a~constrained Hamiltonian system has a momentum section. In Section~\ref{section4}, we discuss a two-dimensional gauged sigma model with boundary and show a boundary term gives a~momentum section. In Section~\ref{section5}, we consider gauging conditions of an $n$-dimensional gauged sigma model with a WZ term and propose a~generalization of a momentum section on a pre-multisymplectic manifold. Section~\ref{section6} is devoted to discussion and outlook.

\section{Momentum section and Hamiltonian Lie algebroid}\label{section2}

In this section, we review a momentum section and a Hamiltonian Lie algebroid introduced in~\cite{Blohmann:2018}.

\subsection{Lie algebroid}\label{section2.1}

A Lie algebroid is a unified structure of a Lie algebra, a Lie algebra action and vector fields on a manifold.
\begin{Definition}\label{Liealgebroid} Let $E$ be a vector bundle over a smooth manifold $M$. A~Lie algebroid $(E, \rho, [-,-])$ is a vector bundle $E$ with a bundle map $\rho\colon E \rightarrow TM$ and a Lie bracket $[-,-]\colon \Gamma(E) \allowbreak \times \Gamma(E) \rightarrow \Gamma(E)$ satisfying the Leibniz rule,
\begin{gather*}
[e_1, fe_2] = f [e_1, e_2] + \rho(e_1) f \cdot e_2,
\end{gather*}
where $e_i \in \Gamma(E)$ and $f \in C^{\infty}(M)$.
\end{Definition}
A bundle map $\rho$ is called an anchor map.
\begin{Example} Let a manifold $M$ be one point $M = \{pt \}$. Then a Lie algebroid is a Lie algebra~$\mathfrak{g}$.
\end{Example}
\begin{Example} If a vector bundle $E$ is a tangent bundle $TM$ and $\rho = \mathrm{id}$, then a bracket $[-,-]$ is a normal Lie bracket of vector fields and $(TM, \mathrm{id}, [-,-])$ is a Lie algebroid.
\end{Example}
\begin{Example} Let $\mathfrak{g}$ be a Lie algebra and assume an infinitesimal action of $\mathfrak{g}$ on a manifold~$M$. The infinitesimal action $\mathfrak{g} \times M \rightarrow M$ determines a map $\rho\colon M \times \mathfrak{g} \rightarrow TM$. The consistency of a~Lie bracket requires a Lie algebroid structure on $(E= M \times \mathfrak{g}, \rho, [-,-])$. This Lie algebroid is called an action Lie algebroid.
\end{Example}

\begin{Example} An important nontrivial Lie algebroid is a Lie algebroid induced from a Poisson structure. A bivector field $\pi \in \Gamma\big({\wedge}^2 TM\big)$ is called a Poisson structure if $[\pi, \pi]_S =0$, where $[-,-]_S$ is a Schouten bracket on $\Gamma(\wedge^{\bullet} TM)$.

Let $(M, \pi)$ be a Poisson manifold. Then, we can define a bundle map, $\pi^{\sharp}\colon T^*M \rightarrow TM$ by $\pi^{\sharp}(\alpha)(\beta) = \pi(\alpha, \beta)$ for all $\beta \in \Omega^1(M)$. A Lie bracket on $\Omega^1(M)$ is defined by the so called Koszul bracket,
\begin{gather*}
[\alpha, \beta]_{\pi} = L_{\pi^{\sharp} (\alpha)}\beta - L_{\pi^{\sharp} (\beta)} \alpha - \md(\pi(\alpha, \beta)),
\end{gather*}
where $\alpha, \beta \in \Omega^1(M)$. Then, $\big(T^*M, \pi^{\sharp}, [-, -]_{\pi}\big)$ is a Lie algebroid.
\end{Example}
One can refer to many other examples, for instance, in \cite{Mackenzie}.

\subsection{Lie algebroid differential}\label{section2.2}
We consider a space of exterior products of sections, $\Gamma(\wedge^{\bullet} E^*)$ on a Lie algebroid~$E$. Its element is called an $E$-differential form. We can define a Lie algebroid differential ${}^E\md\colon \Gamma\big({\wedge}^m E^*\big) \rightarrow \Gamma\big({\wedge}^{m+1} E^*\big)$ such that $\big({}^E\md\big)^2=0$.
\begin{Definition}
A Lie algebroid differential ${}^E\md\colon \Gamma\big({\wedge}^m E^*\big) \rightarrow \Gamma\big({\wedge}^{m+1} E^*\big)$ is defined by
\begin{gather*}
{}^E\md \alpha(e_1, \dots, e_{m+1}) = \sum_{i=1}^{m+1} (-1)^{i-1} \rho(e_i) \alpha(e_1, \dots,
\check{e_i}, \dots, e_{m+1}) \\
\hphantom{{}^E\md \alpha(e_1, \dots, e_{m+1}) =}{}+ \sum_{i, j} (-1)^{i+j} \alpha([e_i, e_j], e_1, \dots, \check{e_i}, \dots, \check{e_j}, \dots, e_{m+1}),
\end{gather*}
where $\alpha \in \Gamma\big({\wedge}^m E^*\big)$ and $e_i \in \Gamma(E)$.
\end{Definition}

It is useful to describe Lie algebroids by means of $\mathbb{Z}$-graded geometry~\cite{Vaintrob}. A graded mani\-fold~$\calM$ with local coordinates $x^i$, $i=1, \dots, \dim M$, and $q^a$, $a=1, \dots, \operatorname{rank} E$, of degree zero and one, respectively, are denoted by $\calM= E[1]$ for some rank~$r$ vector bundle~$E$, where the degree one basis $q^a$ is identified by a section in $E^*$, i.e., we identify $C^{\infty}(E[1]) \simeq \Gamma(\wedge^{\bullet} E)$. A~product for homogeneous elements $f, g \in C^{\infty}(\calM)$ has a~property, $fg = (-1)^{|f||g|} gf$, where $|f|$ is degree of~$f$. Especially, $q^a q^b = - q^b q^a$. We introduce a derivation of degree $-1$, $\frac{\partial}{\partial q^a}$ satisfying $\frac{\partial}{\partial q^a} q^b = \delta^b_a$. Here, a derivation is a linear operator on a space of functions satisfying the Leibniz rule.

The most general degree plus one vector field on $\calM$ has the form
\begin{gather*} %\label{Q}
 Q = \rho_a^i (x) q^a \frac{\partial}{\partial x^i} - \frac{1}{2} C^c_{ab}(x) q^a q^b \frac{\partial}{\partial q^c} ,
\end{gather*}
where $\rho_a^i (x)$ and $C^c_{ab}(x)$ are local functions of~$x$. Since $q^a$ is of degree $1$, $C^c_{ab} = - C^c_{ba}$.

Let $e_a$ be a local basis in $E$ dual to the basis corresponding to the coordinates $q^a$. Two functions $\rho$ and $C$ in $Q$ define a bundle map $\rho\colon E \rightarrow TM$ and a bilinear bracket on $\Gamma(E)$ by means of $\rho(e_a) := \rho_a^i \partial_i$ and $[e_a,e_b]:= C^c_{ab} e_c$. Then, one can prove that these satisfy the definition of a Lie algebroid, iff
 \begin{gather*} Q^2 = 0 .%\label{Q2}
 \end{gather*}
Identifying functions on $C^{\infty}(E[1]) \simeq \Gamma(\wedge^\bullet E^*)$, $Q$ corresponds to a Lie algebroid differential ${}^E \md$. In the remaining of the paper, we identify $C^{\infty}(E[1]) \simeq \Gamma(\wedge^\bullet E^*)$, and $Q$ to ${}^E \md$.

\subsection{Momentum section}\label{section2.3}
In this section, a momentum section on a Lie algebroid $E$ is defined~\cite{Blohmann:2018}. For definition, we suppose a pre-symplectic form $B \in \Omega^2(M)$ on a base manifold $M$, i.e., a closed $2$-form which is not necessarily nondegenerate. A~Lie algebroid $(E, \rho, [-,-])$ is one over a pre-symplectic manifold~$(M, B)$.

We introduce a connection (a linear connection) on $E$, i.e., a covariant derivative $D\colon \Gamma(E) \rightarrow \Gamma(E \otimes T^*M)$, satisfying $D(fe) =f De + \md f \otimes e$ for a section $e \in \Gamma(E)$ and a function $f \in C^{\infty}(M)$. A connection is extended to $\Gamma(M, \wedge^{\bullet} T^*M \otimes E)$ as a degree $1$ operator.

In order to define a momentum section, we consider an $E^*$-valued 1-form $\gamma \in \Omega^1(M, E^*)$ defined by
\begin{gather*}
\bracket{\gamma(v)}{e} = - B(v, \rho(e)),%\label{HH0}
\end{gather*}
where $e \in \Gamma(E)$ and $v \in \mathfrak{X}(M)$ is a vector field. Here $\bracket{-}{-}$ is a natural pairing of~$E$ and~$E^*$. We introduce the following three conditions for a Lie algebroid $E$ on a pre-symplectic mani\-fold~$(M, B)$.
\begin{enumerate}\itemsep=0pt
\item[(H1)] $E$ is \emph{presymplectically anchored with respect to $D$} if
\begin{gather}
D \gamma = 0,\label{HH1}
\end{gather}
where $D$ is a dual connection on $E^*$ defined by
\begin{gather*}
\md \bracket{\mu}{e} = \bracket{D\mu}{e} + \bracket{\mu}{De},
\end{gather*}
for all sections $\mu \in \Gamma(E^*)$ and $e \in \Gamma(E)$. The dual connection extends to a degree 1 operator on $\Omega^k(M, E^*)$.

\item[(H2)] A section $\mu \in \Gamma(E^*)$ is a \emph{$D$-momentum section} if
\begin{gather}
D \mu =\gamma.\label{HH2}
\end{gather}

\item[(H3)] A $D$-momentum section $\mu$ is \emph{bracket-compatible} if
\begin{gather}
{}^E\md \mu (e_1, e_2) = - \bracket{\gamma(\rho(e_1))}{e_2},\label{HH3}
\end{gather}
for all sections $e_1, e_2 \in \Gamma(E)$. We note these conditions have already appeared in~\cite{Kotov:2016lpx} as compatibility conditions of geometric quantities as a metric and a closed differential form with a Lie algebroid structure. The equation (H1) is the same as equation~(6) and (H2) is equation~(7) in~\cite{Kotov:2016lpx}.
\end{enumerate}

A Hamiltonian Lie algebroid is defined as follows.
\begin{Definition}\label{weaklyHamiltonianLA} A Lie algebroid $E$ with a connection $D$ and a section $\mu \in \Gamma(E^*)$ is called \textit{weakly Hamiltonian} if (H1) and (H2) are satisfied. If the condition is satisfied on a neighborhood of every point in~$M$, it is called locally weakly Hamiltonian.
\end{Definition}

\begin{Definition}\label{HamiltonianLA}
A Lie algebroid $E$ with a connection $D$ and a section $\mu \in \Gamma(E^*)$ is called is called \textit{Hamiltonian} if (H1), (H2) and (H3) are satisfied. If the condition is satisfied on a~neighborhood of every point in~$M$, it is called locally Hamiltonian.
\end{Definition}
A bracket-compatible $D$-momentum section, i.e., conditions~(H2) and~(H3) are sufficient in our examples in later section. Thus, in this paper, we mainly consider a bracket compatible momentum section $\mu \in \Gamma(E^*)$ satisfying~(H2) and~(H3), and do not necessarily require that $E$ is presymplectically anchored, i.e.,~(H1). In this case, $D^2 \mu$ is not necessarily zero.

\subsection{Lie algebra case: momentum map}\label{section2.4}
A momentum section is a generalization of a momentum map on a symplectic manifold with a~Lie group action. The definition of a momentum section~(H1), (H2) and (H3) reduces to the definition of a momentum map if a Lie algebroid~$E$ is an action Lie algebroid.

Suppose $B$ is nondegenerate, i.e., $B$ is a symplectic form. Consider an action Lie algebroid on $E = M \times \mathfrak{g}$. It means that an infinitesimal Lie algebra action is given by a bundle map $\rho\colon \mathfrak{g} \times M \rightarrow TM$, such that
\begin{gather*}
[\rho(e_1), \rho(e_2)]= \rho([e_1, e_2]).
\end{gather*}
The bracket in left hand side is a Lie bracket of vector fields. In this case, we can take a~zero connection, $D = \md$. Then, three axioms of a momentum section reduce to the following equations.
\begin{enumerate}\itemsep=0pt
\item[(H1)]
\begin{gather}
 \md \gamma(e) = \md (\iota_{\rho(e)} B) = \calL_{\rho(e)} B = 0. \label{momentmap1}
\end{gather}
This means that $\rho(e)$ is a symplectic vector field.

\item[(H2)] A section $\mu \in \Gamma(M \times \mathfrak{g}^*)$ is regarded as a map $\mu\colon M \rightarrow \mathfrak{g}^*$. Equation~\eqref{HH2} is that a~map~$\mu$ is a Hamiltonian for the vector field $\rho(e)$,
\begin{gather}
\md \mu(e) = \iota_{\rho(e)} B.\label{momentmap2}
\end{gather}
Equation \eqref{momentmap2} leads equation \eqref{momentmap1}.

\item[(H3)] $\md \mu = \gamma$, i.e., $\md \mu = -B(\rho,-)$.

Equation \eqref{HH3} is equivalent to
\begin{gather}
\operatorname{ad}^*_{e_1} \mu(e_2) = \mu([e_1, e_2]).\label{momentmap3}
\end{gather}
for $e_1, e_2 \in \mathfrak{g}$. This means that $\mu$ is $\mathfrak{g}$-equivariant.
\end{enumerate}

Independent conditions are \eqref{momentmap2} and \eqref{momentmap3}, which are the definition of an infinitesimally equivariant momentum map.

Many examples of momentum sections which are not momentum maps have been discussed in~\cite{Blohmann:2018}. One can refer to more examples.

\section{Constrained Hamiltonian system}\label{section3}
We discuss examples of physical systems which have momentum sections and Hamiltonian Lie algebroid structures. In this section, we consider a constrained Hamiltonian mechanics system in $1+0$ dimension analyzed in~\cite{Ikeda:2018rwe}.

Let $(N = T^*M, \omega_{\rm can})$ be a symplectic manifold over a smooth manifold $M$, where $\omega_{\rm can}$ is a canonical symplectic form on~$N$. We take Darboux coordinates $\big(x^i, p_i\big)$ such that $\omega_{\rm can} = \md x^i \wedge \md p_i$. On this symplectic manifold, we consider a dynamical system. Assume a Hamiltonian $H \in C^\infty(N)$, and $r$ constraint functions $\Phi_a = \Phi_a(x, p)$, satisfying the following compatibility condition:

There exist local matrix functions $\lambda_a^b = \lambda_a^b(x, p)$ such that
\begin{gather}\label{Hflow}
\{ H,\Phi_a \} = \lambda^b_{a} \Phi_b ,
\end{gather}
where $\{-, -\}$ is the Poisson bracket induced by the symplectic form $\omega_{\rm can}$. This ensures that $H$ is preserved by the Hamiltonian flow of the constraints on the constraint surface $C := \{\Phi_a = 0 \}$.

Moreover, suppose constraint functions are of the \textit{first class}, i.e., they satisfy{\samepage
\begin{gather} \label{first}
\{ \Phi_a , \Phi_b \} = C^c_{ab} \Phi_c,
\end{gather}
for some functions $C^c_{ab} = C^c_{ab}(x, p)$ on~$N$.}

We assume that constraints $\Phi_a$, $a=1, \dots, r$, are \emph{irreducible}, i.e., $\varphi_C^* ( \md \Phi_1 \wedge \dots \wedge \md \Phi_r )$ is everywhere non-zero, where $\varphi_C \colon C \to N$ is the canonical embedding map of the constraint surface into the original phase space. Moreover, two sets of irreducible constraints $\Phi_a$, $a=1, \dots, r$, and $\tilde \Phi_a$, $a=1, \dots, r$, are \emph{equivalent} if there exist local matrix functions $M_a^b = M_a^b(x, p)$ such that
\begin{gather} \label{equiv}
 \tilde \Phi_a = M^b_a \Phi_b
\end{gather}
 holds true and the matrix $\big(M_a^b\big)_{a,b = 1}^r$ is invertible when restricted to~$C$.

We take setting of the paper~\cite{Ikeda:2018rwe}. We require the canonical symplectic form $\omega_{\rm can} = \md x^i \wedge \md p_i$. Then, there is a natural grading of functions with respect to the monomial degree in the momenta~$p_i$. The space of order~$i$ or less than~$i$ functions is denoted by $C^\infty_{\leq i}(T^*M)$.

As a typical example which appears in physical applications, we consider the case of $\Phi_a \in C^\infty_{\leq 1}(T^*M)$ and $H \in C^\infty_{\leq 2}(T^*M)$. These imply
\begin{gather}\label{affine}
 \Phi_a= \rho_a^i(x) p_i + \alpha_a(x) ,
\end{gather}
and
\begin{gather} \label{H2}
H= \frac{1}{2} g^{ij}(x) p_ip_j+ \beta^i(x)p_i + V(x) .
\end{gather}
Here $\rho_a^i(x)$, $\alpha_a(x)$, $g^{ij}(x)$, $\beta^i(x)$ and $V(x)$ are local functions of~$x$.

We show that this Hamiltonian mechanics system has a momentum section and a Hamiltonian Lie algebroid structure.

\subsection{Lie algebroid structure on constraints}\label{section3.1}

First, we see equation \eqref{first} with \eqref{affine}. As explained in~\cite{Ikeda:2018rwe}, this equation requires an (anchored almost) Lie algebroid structure. Counting an order of $p_i$ in the equivalence condition \eqref{equiv}, matrix functions $M^a_b$ are functions of~$x$. Then, a global structure is a rank~$r$ vector bundle $E$ over~$M$ with transition functions $(M^a_b)_{a,b = 1}^r$.

The Poisson bracket reduces the order by one or less than one since $\{ p_i , x^j \} = \delta_i^j$ and $\{ p_i , p_j \} = 0$. Thus, the equality~\eqref{first} implies~$C^c_{ab} \in C^\infty_0(T^*M) \cong C^\infty(M)$, which is uniquely determined due to the irreducibility condition. The 1st order of~$p$ of equation~\eqref{first} takes the form,
$[\rho_a,\rho_b]^i = C^c_{ab} (x) \rho_c^i$, i.e., globally,
 \begin{gather}
[\rho(e_1), \rho(e_2)] = \rho([e_1, e_2]),\label{LA1}
\end{gather}
for $e_1, e_2 \in \Gamma(E)$.

 Next we apply \eqref{first} to the Jacobi identity $\{\{\Phi_a,\Phi_b\},\Phi_c\} + \operatorname{cycl}(abc)=0$. The 1st order of~$p_i$ gives
 \begin{gather}
\big(C_{ab}^e C_{ce}^d + \partial_j C_{ab}^d \rho_c^j + \operatorname{cycl}(abc) \big) {\rho^i_d} =0 .\label{preJacobi}
\end{gather}
From the irreducibility condition on the constraints and the above identity, we may deduce
\begin{gather*}
 C_{ab}^e C_{ce}^d + \rho_{a}^j \partial_j C_{bc}^d + \operatorname{cycl}(abc) = \sigma^{d}_{abc} ,
\end{gather*}
for some functions $\sigma_{abc}^{d}$ skewsymmetric in the lower indices and $\sigma^{d}_{abc} \rho^i_d=0$. If the anchor map~$\rho$ is assumed injective, we have $\sigma^{d}_{abc} =0$ and
\begin{gather} \label{consist} C_{ab}^e C_{ce}^d + \rho_{a}^j \partial_j C_{bc}^d + \operatorname{cycl}(abc) = 0 .
\end{gather}
It is now straightforward to verify that equations \eqref{LA1} and \eqref{consist} yield Lie algebroid axioms, where the anchor map $\rho\colon E \rightarrow TM$ is defined by $\rho(e_a) = \rho^i_a(x) \partial_i$ and the Lie bracket is defined by $[e_a, e_b] = C_{ab}^c(x) e_c$ for a basis $e_a$ of the fiber of~$E$. We remark that the equivalence \eqref{equiv} takes care of the equivalence of the two sides to not depend on the choice of a chosen frame.

A vector bundle with a bundle map $\rho\colon E \rightarrow TM$ and a bilinear bracket $[-, -]$ is an anchored almost Lie algebroid $(E, \rho, [-,-])$ if $\rho$ and $[-,-]$ satisfy
\begin{gather*}
[\rho(e_1),\rho(e_2)] = \rho([e_1, e_2]),\\ %\label{anchoredalmost1}
[e_1, f e_2] = f [e_1, e_2] + \rho(e_1)f \cdot e_2.%\label{anchoredalmost2}
\end{gather*}
If $\rho$ is not injective, we can take an anchored almost Lie algebroid since it satisfies equation~\eqref{preJacobi}. We do not discuss these cases in this paper.

We can take a more general algebroid satisfying $\sigma^{d}_{abc} \rho^i_d=0$ such as a Courant algebroid. We leave such cases to other analysis.

The second term $\alpha_a$ in $\Phi_a$ is considered as components of an $E$-1-form, $\alpha= \alpha_a(x) e^a \in \Gamma(E^*)$, where $e^a$ is a basis on $E^*$. The Poisson bracket \eqref{first} is equivalent to the condition on $\alpha$,
\begin{gather}\label{Edalpha1}
{}^E\md \alpha = 0 .
\end{gather}
On the other hand, $\alpha$ is determined by \eqref{affine} only up to additions of the form $\alpha_a \mapsto \alpha_a + \rho_a^i(x) \partial_i f(x)$, for a function $f$ on $M$, which does not modify the symplectic form. Since such additions to $\alpha$ are the ${}^E\md$-exact ones, we see that 0th order deformations of $p$ in first class constraints \eqref{affine} are parametrized by the Q-cohomology of the Lie algebroid at degree one,
\begin{gather*}%\label{coho}
 [\alpha] \in H_Q^1(E[1]) .
\end{gather*}

Equation \eqref{first} and injective assumption for $\rho$ gives a Lie algebroid structure on~$E$ and equation~\eqref{Edalpha1}.

\subsection{Hamiltonian, metric and connection}\label{section3.2}

In this section, we explain geometric structures induced from the Hamiltonian \eqref{H2} and the Poisson bracket~\eqref{Hflow} discussed in~\cite{Ikeda:2018rwe}. Suppose that in \eqref{H2} the symmetric matrix~$g^{ij}$ has an inverse. Then a symmetric tensor~$g^{ij}$ corresponds to an inverse of a metric~$g$ on~$M$. $\beta := \beta^i \partial_i$ is a~vector field and~$V(x)$ is a~global potential function on~$M$. Counting order of $p$ in equation~\eqref{Hflow}, $\lambda_a^b$~is a 1st order function of~$p$, thus it is assumed that $\lambda_a^b = g^{ij}(x) \Gamma_{aj}^{b}(x) p_i + \tau_a^{b}(x)$. From consistency of equation~\eqref{Hflow} with transition functions $M^a_b$ given by equivalence of~$\Phi_a$, we obtain
 \begin{gather}
 \Gamma^{\prime} = M \Gamma M^{-1} + \md M M^{-1},\nonumber \\ %\label{transformationofGamma}\\
 \tau^{\prime} = M \tau M^{-1} + \big(\beta, \md M M^{-1}\big),\label{transformationoftau}
 \end{gather}
where $\Gamma_a^b = \Gamma_{aj}^{b}\md x^j$. $\Gamma_a^b$ transforms as a connection 1-form on~$E$. $\tau_a^b$ is a local matrix transforming in equation~\eqref{transformationoftau}.\footnote{An interesting $\tau$ is $\tau = (\Gamma, \beta)$, which case is analyzed later in this section.}

We can absorb the term linear in the momenta in the Hamiltonian, $\beta^i \mapsto 0$, at the expense of redefining the potential $V$ and the $E$-1-forms $\alpha$ and simultaneously twisting the symplectic form $\omega_{\rm can}$ by a magnetic field $B=\md A \in \Omega^2(M)$ as
\begin{gather*} \omega = \omega_{\rm can} + B , %\label{B}
\end{gather*}
where $A_i = g_{ij} \beta^j$ and $A=A_i(x) \md x^i$. The globally defined $2$-form $B = \md A$ is obviously regarded as a pre-symplectic form since $\md B=0$.

By the above redefinition, constraints and the Hamiltonian become
 \begin{gather*} %\label{affine2}
 \Phi_a^{\prime}= \rho_a^i(x) p_i + \alpha^{\prime}_a(x) ,\\
%\label{H22}
H = \frac{1}{2} g^{ij}(x) p_ip_j + V^{\prime}(x) .
 \end{gather*}
Here, $\alpha^{\prime}$ is an E-1-form defined by $\bracket{\alpha^{\prime}}{e} = \bracket{\alpha}{e} - \iota_{\rho(e)} A$ for all $e \in \Gamma(E)$, and $V^{\prime}$ is defined by $V^{\prime}(x) = V(x) - \frac{1}{2} g(\beta, \beta)$. Equations \eqref{first} and \eqref{Hflow} change but are similar equations,
 \begin{gather} \label{first2}
\{\Phi^{\prime}_a, \Phi^{\prime}_b\} = C_{ab}^c \Phi^{\prime}_c,\\
 \label{Hflow2}
\{H, \Phi^{\prime}_a\} = \lambda_a^{\prime b} \Phi^{\prime}_b,
 \end{gather}
where $\tau^{\prime} = \tau - g^{-1}(\Gamma, A)$ and $\lambda^{\prime} = \lambda - g^{-1}(\Gamma, A) = g^{-1}(\Gamma, p) + \tau^{\prime}$.

After the above redefinition, we show that geometric structure described by equations~\eqref{first2} and~\eqref{Hflow2} have a structure of a momentum section.

The 1st order term of $p$ in equation~\eqref{first2} gives the same conditions as~\eqref{first}, i.e.,~\eqref{first2} requires a Lie algebroid structure on the vector bundle~$E$ with the same anchor map $\rho$ and Lie bracket $[-,-]$ before the redefinition. In the 0th order term of~$p$ in equation~\eqref{first2}, the affine constraints~$\alpha$ changes to
\begin{gather}
{}^E \md \alpha^{\prime} = - \rho^*(B) ,\label{Edalpha2}
\end{gather}
since the new symplectic form $\omega$ gives the Poisson bracket $\{ p_i,p_j\} = B_{ij}$. Here $\rho^*$ is the induced map of the anchor to $\Omega^{\bullet}(M)$, mapping ordinary differential forms to $E$-differential forms. In particular, $\rho^*(B)= \frac{1}{2} B_{ij} \rho_a^i \rho_b^j q^a q^b \in \Gamma\big({\wedge}^2E^*\big)$. Equation~\eqref{Edalpha2} is the same as equation~\eqref{HH3} in the condition~(H3) by identifying $\mu = \alpha^{\prime}$.

Let us analyze equation \eqref{Hflow2}. As already pointed, the transformation property of $\Gamma^a_{bi}$ under the transition function $M^a_b$ shows $\Gamma^a_{bi}$ is a connection 1-form, thus this defines a Lie algebroid connection $D\colon \Gamma(E) \rightarrow \Gamma(E \otimes T^*M)$. $D$ and $\rho$ can be combined to define an $E$-connection ${}^E\nabla\colon \Gamma(TM) \rightarrow \Gamma(TM \otimes E^*)$ on $TM$:
\begin{gather}
{}^E\nabla_{\! e} v := \calL_{\rho(e)} v + \rho(D_v e),
\label{Econnection}
\end{gather}
where $v \in \mathfrak{X}(M)$ and $e \in \Gamma(E)$. An $E$-connection is extended to a tensor product space of $TM$ and $T^*M$.

Equation \eqref{Hflow2} then gives three conditions by considering it to 2nd, 1st, and 0th order in the momenta. To 2nd order, we obtain the geometrical compatibility equation,
\begin{gather*} {}^E\nabla g = 0 ,%\label{ginv'}
\end{gather*}
on the metric $g$.

To 1st order, we get another condition on the system of constraints, It relates the exterior covariant derivative of $\alpha^{\prime}$ induced by~$D$, $D \alpha^{\prime} \in \Gamma(E^* \otimes T^*M)$, to the anchor map $\rho$, now regarded as a section of~$E^* \otimes TM$:
\begin{gather}
 D \alpha^{\prime} = \gamma + (\tau^{\prime t} \otimes g_{\flat}) \rho,\label{dalpha}
\end{gather}
where $\gamma \in \Omega^1(M, E^*)$ is a 1-form taking a value on $E^*$ appeared in the definition of a momentum section, $\tau^{\prime t} \colon E^* \to E ^*$, the transposed of $\tau^{\prime}$, and $g_\flat \colon TM \to T^*M$, $v \mapsto \iota_vg$, as maps on the corresponding sections. To 0th order one finds that the potential $V^{\prime}$ has to satisfy
\begin{gather*} {}^E\md V^{\prime} = \tau^{\prime}(\alpha^{\prime}) . %\label{dV}
\end{gather*}

If $\tau^{\prime} =0$, equation \eqref{dalpha} becomes
\begin{gather*}
 D \alpha^{\prime} = \gamma,%\label{dalpha2}
\end{gather*}
which is the condition (H2), i.e., equation \eqref{HH2}, since $\mu = \alpha^{\prime}$. The condition $\tau^{\prime} =0$ is $\tau = g(\Gamma, A)$. The remaining condition of a momentum section is the condition~(H1), i.e., equation~\eqref{HH1}, $D\gamma=0$. In general, this constrained Hamiltonian system does not satisfy~\eqref{HH1}.\footnote{Using an $E$-connection \eqref{Econnection} and its extension to $\Omega^2(M)$, equation \eqref{HH1} is equivalent to ${}^E\nabla B =0$, where $\gamma_{ia} = - B_{ij} \rho^j_a$ and
${}^E\nabla_a B_{ij} = \rho_a^k \partial_k B_{ij} + \partial_i \rho_a^k B_{kj} + \partial_j \rho_a^k B_{ik} + \rho^k_b \Gamma_{aj}^b B_{ik} + \rho^k_b \Gamma_{ai}^b B_{kj} =0$.}

Therefore, we obtain the following result:
\begin{Theorem}We consider the constraint Hamiltonian system satisfying equations~\eqref{Hflow} and \eqref{first} with constraints~\eqref{affine} and a Hamiltonian~\eqref{H2}. Then, $B= \md (g(\beta,-))$ is a pre-symplectic form. If $\rho$ is injective and $\tau^{\prime} = \tau - g(\Gamma, A) = \tau - (\Gamma, \beta) = 0$, $\alpha^{\prime} = \alpha - \iota_{\rho} A$ is a bracket compatible $D$-momentum section on a Lie algebroid $E$ with respect to a connection~$D$ defined by a~connection $1$-form~$\Gamma^b_{a}$. Moreover, if $D \gamma = D(- \iota_{\rho} B) =0$, it is pre-symplectically anchored.
\end{Theorem}
In $\tau^{\prime} \neq 0$ case, this constrained Hamiltonian system gives a generalization of a momentum section. It is interesting to see this generalization in a future work.

\section{Two-dimensional sigma model with boundary}\label{section4}
In this section, we analyze a momentum section and a Hamiltonian Lie algebroid structure in a two-dimensional sigma model. If a base manifold is in two dimensions and with boundary, a~momentum section naturally appears.

Let $\Sigma$ be a two-dimensional manifold and $M$ be a $d$-dimensional target manifold. $X\colon \Sigma \rightarrow M$ is a smooth map from~$\Sigma$ to~$M$. We start at the following sigma model action with a $2$-form B-field,
\begin{gather}
S = \frac{1}{2} \int_{\Sigma} g_{ij}(X) \md X^i \wedge *\md X^j + b_{ij}(X) \md X^i \wedge \md X^j,\label{sigmamodelwithB}
\end{gather}
where $g$ is a metric and $b \in \Omega^2(M)$ is a closed $2$-form on~$M$. $g_{ij}(X)$ and $b_{ij}(X)$ are their pullbacks to $\Sigma$. This action is invariant under diffeomorphisms on a worldsheet~$\Sigma$ and on a~target space~$M$.

We analyze a general condition that the action $S$ is invariant under other symmetries on~$M$. In a general setting, an element of a vector space~$V$, or more generally, a section of the vector bundle~$E$ on~$M$, $e \in \Gamma(E)$ acts on $M$ as an infinitesimal transformation generated by a~vector field. A~transformation is determined by defining a bundle map to a tangent bundle, \mbox{$\rho\colon E \rightarrow TM$}. Suppose that $\rho$ define an infinitesimal gauge transformation of~$X$ as
\begin{gather}
\delta X^i = \rho(\epsilon)^i = \rho^i_a(X) \epsilon^a,\label{transformationofX}
\end{gather}
where $i = 1, 2, \dots, d$ are indices of local coordinates on $M$, $\epsilon \in \Gamma(X^*E)$ is a parameter (a gauge parameter), and $\rho(e_a) = \rho^i_a(X) \partial_i$ by taking a basis of~$E$,~$e_a$.

By straight computations, the action \eqref{sigmamodelwithB} is in invariant under the transformation \eqref{transformationofX}, iff
\begin{gather}
\calL_{\rho(e_a)} g = 0,\label{killingg}\\
\calL_{\rho(e_a)} b =\md \beta_a,\label{killlingB}\\
[\rho(e_a), \rho(e_b)] = \rho([e_a, e_b]),\label{almostLiealgebroid}
\end{gather}
where $\calL$ is a Lie derivative and $\beta_a \in \Omega^1(M, E^*)$ is a 1-form taking a value on~$E^*$. A vector field $\rho(e_a)$ satisfying equation \eqref{killingg} is called a Killing vector field. From equation~\eqref{almostLiealgebroid}, a vector bundle is an anchored almost Lie algebroid.

In this paper, $E$ is a Lie algebroid. In this case, the action~$S$ is invariant if equations~\eqref{killingg} and~\eqref{killlingB} are satisfied.

\subsection{Gauged sigma model}\label{section4.1}
We can generalize the above theories by gauging the action~\eqref{sigmamodelwithB}. `Gauging' is a deformation of the action using a connection 1-form $A \in \Omega^1(\Sigma, X^*E)$.

A pullback of a basis of a 1-form on $M$, $\md X^i$, is `gauged' using a covariant derivative with respect to a~connection~$A$ as
\begin{gather*}
F^i = D X^i = \md X^i - \rho^i_a(X) A^a.
\end{gather*}
We can assume $A^a$ has a genuine infinitesimal gauge transformation,
\begin{gather*}
\delta A^a = \md \epsilon^a + [A, \epsilon]^a = \md \epsilon^a + C_{bc}^a A^b \epsilon^c,
\end{gather*}
however, $C_{bc}^a = C_{bc}^a(X)$ is not necessarily constant but a local function on $M$. We consider a target space covariant version of the gauge transformation by introducing (a pullback of) a~connection on $M$, $\Gamma_{bi}^a(X)$:\footnote{We can consider a more general ansatz of a gauge transformation as $\delta A^a = D \epsilon^a + [A,\epsilon]^a= \md \epsilon^a + C_{bc}^a A^b \epsilon^c + \Delta A^a$, where $\Delta A^a$ is a 1-form taking a value on a pullback of $E$, which is linear with respect to the infinitesimal parameter~$\epsilon^a$ \cite{ Chatzistavrakidis:2016jci, Chatzistavrakidis:2017tpk, Chatzistavrakidis:2016jfz}.}
\begin{gather*}
\delta A^a = \md \epsilon^a + C_{bc}^a(X) A^b \epsilon^c + \Gamma_{bi}^a(X) \epsilon^b DX^i,
\end{gather*}
where {the gauge transformation is covariant} under the target space diffeomorphism. In summary, we choose gauge transformations,
\begin{gather}
\delta X^i = \rho^i_a(X) \epsilon^a,\label{gaugetransformationX}\\
\delta A^a =\md \epsilon^a + C_{bc}^a(X) A^b \epsilon^c + \Gamma_{bi}^a(X) \epsilon^b DX^i.\label{gaugetransformationA}
\end{gather}
The action \eqref{sigmamodelwithB} is generalized to a gauged sigma model action by `gauging' the symmetry to infinitesimal transformations~\eqref{gaugetransformationX} and~\eqref{gaugetransformationA}. Since the manifold $\Sigma$ has boundary, we take the following ansatz for a gauged sigma model action:
\begin{gather}
S = \frac{1}{2} \int_{\Sigma} g_{ij}(X) DX^i \wedge *DX^j+ b_{ij}(X) \md X^i \wedge \md X^j + \int_{\partial \Sigma} \eta_i(X) \md X^i + \mu_a (X) A^a,\label{gaugedsigmamodelwithB}
\end{gather}
where the last two terms are the most general possible boundary terms with some arbitrary local functions $\eta_i(X)$ and~$\mu_a (X)$. $\eta_i(X)dX^i$ is a pullback of a 1-form on a target space~$M$ and~$\mu_a (X)$ is a pullback of an element $\Gamma(E^*)$ on a target space~$M$. Requiring~\eqref{gaugedsigmamodelwithB} is invariant under gauge transformations~\eqref{gaugetransformationX} and \eqref{gaugetransformationA}, we obtain geometric conditions for a~metric~$g$, a $2$-form~$b$ and~$\rho$ and a bracket~$[-,-]$. We obtain the following conditions for the metric, $\rho$ and a bracket,
\begin{gather}
 \calL_{\rho(e_a)} g= \Gamma_a^b \vee \iota_{\rho(e_b)} g,\label{conditionofgsm1}\\
[\rho(e_a), \rho(e_b)] = \rho([e_a, e_b]),\label{conditionofgsm3}
\end{gather}
where $\vee$ is a symmetric product of 1-forms. Equation \eqref{conditionofgsm1} is equivalent to ${}^E\nabla g=0$. Though equation~\eqref{conditionofgsm3} is satisfied if $(E, \rho, [-,-])$ is an anchored almost Lie algebroid, we suppose $(E, \rho, [-,-])$ is a true Lie algebroid now.

Next we analyze a condition for a $2$-form B-field $b$. Using $\md b=0$, the gauge transformation for $S_b = \frac{1}{2} \int_{\Sigma} b_{ij}(X) \md X^i \wedge \md X^j$ is
\begin{gather*}
\delta S_b = \int_{\Sigma} \calL_{\rho(\epsilon)} b= \int_{\Sigma} \md \iota_{\rho(\epsilon)} b = \int_{\partial \Sigma} \iota_{\rho(\epsilon)} b. %\label{gaugetransfofB}
\end{gather*}
Thus, requirement of gauge invariance of the total action $\delta S=0$ gives the conditions including quantities of boundary terms. In local coordinates, straight computations give three equations,
\begin{gather}
 \mu_a = - \eta_i \rho^i_a,\label{conditionof2d1}\\
\rho^j_a b_{ji} + \rho^j_a \partial_j \eta_i + \eta_j \partial_i \rho^j_a+ \Gamma^b_{ai} \mu_b = 0,\label{conditionof2d2}\\
\rho^i_a \partial_i \mu_b - C^c_{ab} \mu_c - \rho^i_b \Gamma^c_{ai} \mu_c = 0.\label{conditionof2d3}
\end{gather}
The first condition \eqref{conditionof2d1} is $\mu(e) = - \iota_{\rho(e)} \eta$ for $e \in \Gamma(E)$, the second and third conditions~\eqref{conditionof2d2} and~\eqref{conditionof2d3} are equivalent to (H2) and (H3), where we identify $B = b + \md\eta$. Thus, we obtain the following result.
\begin{Theorem}We consider a gauged sigma model with boundary, \eqref{gaugedsigmamodelwithB}. Let $\mu(e) = - \iota_{\rho(e)} \eta \in \Gamma(E^*)$ and $B = b + \md\eta \in \Omega^2(M)$. Let $D$ be a connection defined by~$\Gamma_a^b$. Then, $\mu$ is a bracket compatible $D$-momentum section with respect to the connection~$D$ with a pre-symplectic form~$B$. If~$B$ satisfies~(H1), it is pre-symplectically anchored.
\end{Theorem}

Finally, we comment a gauge algebra generated by gauge transformations \eqref{gaugetransformationX} and \eqref{gaugetransformationA}. Gauge transformations must consist of a closed algebra at least on an orbit of equations of motion. Closure conditions of a gauge algebra generated $[\delta_1, \delta_2] \sim \delta_3$ by equations~\eqref{gaugetransformationX} and~\eqref{gaugetransformationA} impose extra equations for~$X^i$ and~$A^a$, which are topological conditions.

This gives topological condition on a external gauge field $A^a$ and geometry of $M$. One can refer to \cite{Bouwknegt:2017xfi} and \cite{Wright:2019pru}
for more analysis related T-duality. See also~\cite{Higgins}.

\section{Momentum section on pre-multisymplectic manifold}\label{section5}
In this section, we propose a generalization of a momentum section to a pre-multisymplectic mani\-fold. Our strategy is to generalize a gauged sigma model in Section~\ref{section4}. We generalize a~$2$-form B-field~$b$ to a higher $(n+1)$-form~$h$ and a~two-dimensional manifold $\Sigma$ to a higher-dimensional manifold.
We naturally obtain a generalization of a momentum section from consistency of these gauged sigma models.

\subsection[Gauged sigma model in $n$ dimensions with Wess--Zumino term]{Gauged sigma model in $\boldsymbol{n}$ dimensions with Wess--Zumino term}\label{section5.1}
We introduce a pre-$n$-plectic manifold.
\begin{Definition}
A pre-$n$-plectic manifold is $(M, h)$, where $M$ is a smooth manifold and~$h$ is a~closed $(n+1)$-form on $M$.
\end{Definition}
A pre-$n$-plectic manifold is also called a pre-multisympletic manifold for $n \geq 2$. A pre-$n$-plectic manifold is called an $n$-plectic manifold if~$h$ is nondegenerate, i.e., if $\iota_{v} h =0$ for a vector field $v \in \mathfrak{X}(M)$ is equivalent to $X=0$.

We also introduce a metric $g$ on $M$. Moreover, by introducing an $(n+1)$-dimensional mani\-fold~$\Xi$ with boundary $\Sigma = \partial \Xi$, and a map $X\colon \Sigma \rightarrow M$, we consider a nonlinear sigma model with a Wess--Zumino term. Compatibility conditions of $h$ and $g$ are determined from gauge invariance of the following sigma model action with a Wess--Zumino term:
\begin{gather}
S = \int_{\Sigma} \frac{1}{2} g_{ij}(X) \md X^i \wedge *\md X^j+ \int_{\Xi} \frac{1}{(n+1)!} h_{i_1\dots i_{n+1}}(X) \md X^{i_1} \wedge \dots \wedge \md X^{i_{n+1}},\label{ndsigmamodel}
\end{gather}
where $g(X)$ is a pullback of a metric $g$ and $h(X) = \frac{1}{(n+1)!} h_{i_1\dots i_{n+1}}(X) \md X^{i_1} \wedge \dots \wedge \md X^{i_{n+1}}$ in the second term is a pullback of a closed $(n+1)$-form $h$ on~$M$. The $n=2$ case is most important since this is string sigma model with an NS-flux $3$-form~$H$.

Invariance conditions of $S$ under the transformation~\eqref{transformationofX} of~$X$ as in Section~\ref{section4} gives a similar condition,
\begin{gather}
\calL_{\rho(e_a)} g = 0,\nonumber\\ %\label{killingg2}\\
\calL_{\rho(e_a)} h =\md \beta_a,\nonumber\\ %\label{killlingB2}\\
[\rho(e_a), \rho(e_b)] = \rho([e_a, e_b]),\label{almostLiealgebroid2}
\end{gather}
where $\beta$ is an $n$-form taking a value on $E^*$. Equation~\eqref{almostLiealgebroid2} require an anchored almost Lie algebroid structure on a target vector bundle~$E$.

We consider gauging of an $n$-dimensional sigma model \eqref{ndsigmamodel} by introducing a connection $A \in \Omega^1(\Sigma, X^*E)$ and gauge transformations~\eqref{gaugetransformationX} and~\eqref{gaugetransformationA}. Here, we consider the case that a~vector bundle $E$ is a Lie algebroid for~\eqref{almostLiealgebroid2} again. We take a~Hull--Spence type ansatz~\cite{Hull:1990ms} for a~gauged action, but in our case a~gauge structure is not a Lie algebra but a Lie algebroid. The ansatz is
\begin{gather}
S = S_g + S_h + S_{\eta},\label{ndgaugedsigmamodel}
\end{gather}
where
\begin{gather*}
S_g = \int_{\Sigma} \frac{1}{2} g_{ij} DX^i \wedge *DX^j,\\
S_h = \int_{\Xi} \frac{1}{(n+1)!} h_{i_1\dots i_{n+1}}(X) \md X^{i_1} \wedge \dots \wedge \md X^{i_{n+1}},\\
S_{\eta} = \int_{\Sigma} \sum_{k=0}^n \frac{1}{k!(n-k)!} \eta^{(k)}_{i_1 \dots i_k a_{k+1} \dots a_{n}} (X) \md X^{i_1} \wedge \dots \wedge \md X^{i_k}
\wedge A^{a_{k+1}} \wedge \dots \wedge A^{a_{n}},
\end{gather*}
where $\eta^{(k)}$ is a pullback of a $k$-form on $M$ taking a value on $\wedge^{n{-}k}\!E^*$, i.e., $\eta^{(k)} \!\in\! X^* \Omega^k(M, \wedge^{n{-}k}\!E^*)$.

We require gauge invariance of the above gauged action under the gauge transformations~\eqref{gaugetransformationX} and \eqref{gaugetransformationA},
which are the same ones as in two-dimensional case Section~\ref{section4}. Requirement of gauge invariance imposes conditions for pullbacks of coefficient functions $g \in \Gamma\big(S^2 T^*M\big)$, $h \in \Omega^{n+1}(M)$ and $\eta^{(k)} \in \Omega^k\big(M, \wedge^{n-k}E^*\big)$. These identities gives geometric identities of a metric~$g$,~$H$ and $\eta^{(k)}$ on the vector bundle $E$ on $M$ before pullbacks.\footnote{We use the same notation for geometric quantities on~$M$ and their pullbacks. We propose that this structure gives a momentum section in a pre-$n$-plectic manifold.}

From concrete computations, the condition of $g$ is
\begin{gather*}
 \calL_{\rho(e_a)} g= \Gamma_a^b \vee \iota_{\rho(e_b)} g,%\label{conditionofgsm11}
\end{gather*}
as in the case of the two-dimensional sigma model. For $h$ and $\eta^{(k)}$ on~$M$, we obtain the following conditions on~$M$:\footnote{Note that we obtain identities on $h \in \Omega^{n+1}(M)$ and $\eta^{(k)} \in \Omega^k\big(M, \wedge^{n-k}E^*\big)$ from conditions for their pullbacks in the gauged sigma model~\eqref{ndgaugedsigmamodel}.} two algebraic conditions,
\begin{gather}
 \eta^{(k-1)}(e_k, \dots, e_n) = (-1)^k \iota_{\rho(e_k)}\eta^{(k)}(e_{k+1}, \dots, e_n) + \operatorname{cycl}(e_k, \dots, e_n),\label{conditionofgsm21}\\
 \iota_{\rho(e_k)} \eta^{(k)}(e_{k+1}, \dots, e_{k+m}, \dots, e_n)+ \iota_{\rho(e_{k+m})} \eta^{(k)}(e_{k+1}, \dots, e_k, \dots, e_n) =0,\nonumber \\
\qquad k=1, \dots, n-1, \quad m=1, \dots, n-k,\label{conditionofgsm22}
\end{gather}
and three differential equations,
\begin{gather}
D \eta^{(n-1)}(e) = \iota_{\rho(e)} \widetilde{h}, \qquad k=n,\label{conditionofgsm23}\\
\calL_\rho(e) \eta^{(k)} (e_{k+1}, \dots, e_n)+ \sum_{i=1}^{n-k} (-1)^i \eta^{(k)} ([e, e_{k+i}], e_{k+1}, \dots,\check{e}_{k+i}, \dots, e_n)\nonumber \\
\qquad{} + \sum_{i=1}^{n-k} (-1)^i \langle \Gamma, \rho (e) \rangle \wedge \eta^{(k)} (e_{k+1}, \dots, e_n)\nonumber\\
\qquad{} - \sum_{i=1}^{n-k} (-1)^i \Gamma(e)
\wedge \iota_{\rho(e_{k+i})} \eta^{(k)} (e_{k+1}, \dots,\check{e}_{k+i}, \dots, e_n)\nonumber \\
\qquad {} + \sum_{i=1}^{n-k} (-1)^i \langle \iota_{\rho(e_{k+i})} \Gamma(e)\stackrel{\wedge}{,} \eta^{(k)} (e_{k+1}, \dots,\check{e}_{k+i}, \dots, e_n) \rangle
=0,\qquad k=1, \dots, n-1,\!\!\!\label{conditionofgsm24}\\
\calL_\rho(e) \eta^{(0)} (e_{1}, \dots, e_n)+ \sum_{i=1}^{n} (-1)^i \eta^{(0)} ([e, e_{k+i}], e_{k+1}, \dots,\check{e}_{k+i}, \dots, e_n)\nonumber \\
\qquad{}+ \sum_{i=1}^{n} (-1)^i \langle \iota_{\rho(e_{i})} \Gamma(e) \stackrel{\wedge}{,} \eta^{(0)} (e_{1}, \dots,\check{e}_{i}, \dots, e_n) \rangle=0,\qquad k=0,
\label{conditionofgsm25}
\end{gather}
where $\widetilde{h} = h + \md \eta^{(n)}$, $e, e_i \in \Gamma(E)$, $i = k, \dots, n$, $\Gamma$ is a~connection 1-form on~$E$, and $\bracket{-}{-}$ is a~natural pairing of~$E^*$ and~$E$. Notation $\stackrel{\wedge}{,}$ means both a wedge product on~$\Omega^k(M)$ and a pairing of $E$ and $E^*$. Note that $\delta S_h = \int_{\Xi} \calL_{\rho(\epsilon)} h = \int_{\Sigma} \iota_{\rho(\epsilon)} h$ since $\md h=0$. For $k=n-1$, equation \eqref{conditionofgsm24} is also written as
\begin{gather*}
{}^E\md \eta^{(n-1)} (e_1, e_2) - D \eta^{(n-2)} (e_1, e_2) =0.%\label{conditionofgsm242}
\end{gather*}

In $n=1$, equations \eqref{conditionofgsm21}--\eqref{conditionofgsm25} reduce to conditions of a momentum section (H2) and (H3) by setting $\mu = \eta^{(0)}$, $\gamma = \eta^{(1)}$ and $B = \widetilde{h}$. In $n=2$, equations \eqref{conditionofgsm21}--\eqref{conditionofgsm25} give gauging conditions of target geometry in \cite{Chatzistavrakidis:2016jci}.

It is natural to impose the following condition corresponding to the condition (H1),
\begin{gather}
D \iota_{\rho} \widetilde{h} =0.\label{conditionofgsm26}
\end{gather}
However, this condition is not needed for gauge invariance of a gauged sigma model. As a result, we need not impose this condition on the definition of a momentum section.

Finally, we obtain the following definition of a multimomentum section on a pre-mutli\-sym\-plec\-tic manifold.

\begin{Definition}Let $(M, \widetilde{h})$ be a pre-$n$-plectic manifold, where $\widetilde{h}$ is a closed $(n+1)$-form, and $(E, \rho, [-,-])$ be a Lie algebroid over $M$. We define the following three conditions corresponding to (H1), (H2) and (H3).
\begin{description}\itemsep=0pt
\item[{\rm (HM1)}] $E$ is a \emph{pre-$n$-plectically anchored with respect to $D$} if
\begin{gather*}
D \gamma = 0,%\label{HHM1}
\end{gather*}
where $\gamma = \iota_{\rho} \widetilde{h} \in \Omega^n(M, E^*)$.

\item[{\rm (HM2)}] $\eta^{(n-1)} \in \Omega^{n-1}(M, E^*)$ is a \emph{$D$-multimomentum $(D$-momentum$)$ section} if it satisfies equation~\eqref{conditionofgsm23},
\begin{gather*}
D \eta^{(n-1)}(e) = \iota_{\rho(e)} \widetilde{h}.
\end{gather*}

\item[{\rm (HM3)}] We define a descent set of multimomentum sections $\big(\eta^{(k)}\big)_{k=0}^{n-2}$ by equations~\eqref{conditionofgsm21} and \eqref{conditionofgsm22}, where $\eta^{(k)} \in \Omega^{k}\big(M, \wedge^{n-k}E^*\big)$. A $D$-multimomentum section and its descents $\big(\eta^{(k)}\big)_{k=0}^{n-1}$ are \emph{bracket-compatible} if~\eqref{conditionofgsm24} and~\eqref{conditionofgsm25} are satisfied.
\end{description}
\end{Definition}

Under this definition, we have the same definition of a weakly Hamiltonian Lie algebroid, Definition~\ref{weaklyHamiltonianLA}, and a Hamiltonian Lie algebroid, Definition~\ref{HamiltonianLA}, but a~momentum section is a~set of multimomentum sections $\eta^{(k)}$ on a pre-multisymplectic manifold $\big(M, \widetilde{h}\big)$.
A Hamiltonian Lie algebroid on a pre-multisymplectic manifold is defined as follows.
\begin{Definition}\label{weaklyHamiltonianLAM} A Lie algebroid $E$ with a connection $D$ and a section $\eta^{(n-1)} \in \Omega^{n-1}(M, E^*)$ is called \textit{weakly Hamiltonian} if (HM1) and (HM2) are satisfied. If the condition is satisfied on a~neighborhood of every point in~$M$, it is called locally weakly Hamiltonian.
\end{Definition}

\begin{Definition}\label{HamiltonianLAM} A Lie algebroid $E$ with a connection $D$ and a section $\eta^{(k)} \in \Omega^k\big(M, \wedge^{n-k}E^*\big)$, $k=0, \dots, n-1$ is called \textit{Hamiltonian} if (HM1), (HM2) and (HM3) are satisfied. If the condition is satisfied on a neighborhood of every point in $M$, it is called locally Hamiltonian.
\end{Definition}

We summarize a geometric structure of a gauge sigma model with a $(n+1)$-form flux $h$ using the terminology of multimomentum sections.
\begin{Theorem}We consider an $n$-dimensional gauged sigma model with WZ term~\eqref{ndgaugedsigmamodel}. Then, $\eta^{(k)} \in \Omega^k\big(M, \wedge^{n-k}E^*\big)$, $k=0, \dots, n-1$, are a bracket compatible $D$-multimomentum section and descents with a pre-$n$-plectic form $\widetilde{h} = h + \md\eta^{(n)}$. If $\widetilde{h}$ satisfies~{\rm (HM1)}, it is pre-$n$-plectically anchored.
\end{Theorem}

\subsection{Momentum map on multisymplectic manifold: Lie algebra case}\label{section5.2}
Let a Lie algebroid be an action Lie algebroid $E = M \times \mathfrak{g}$. Then, we can take a trivial connection $\md=D$, and a momentum section on a pre-$n$-plectic manifold reduces to a (multi)momentum map on a pre-symplectic manifold.

Conditions \eqref{conditionofgsm21}--\eqref{conditionofgsm25} reduce to
\begin{gather}
\eta^{(k-1)}(e_k, \dots, e_n) = (-1)^k\operatorname{ad}^*_{e_k} \eta^{(k)}(e_{k+1}, \dots, e_n) + \operatorname{cycl}(e_k, \dots, e_n),\nonumber\\ %\label{conditionofgsm31}\\
\operatorname{ad}^*_{e_k}\eta^{(k)}(e_{k+m}, \dots, e_{k+m}, \dots, e_n)+\operatorname{ad}^*_{e_{k+1}}\eta^{(k)}(e_{k+1}, \dots, e_k, \dots, e_n) =0,\nonumber \\
\qquad k=1, \dots, n-1, \quad m=1, \dots, n-k, \nonumber\\ %\label{conditionofgsm32}\\
\md \eta^{(n-1)} = \iota_{\rho_a} \widetilde{h}, \qquad k=n, \label{conditionofgsm33}\\
\md \eta^{(k-1)}(e, e_{k+1}, \dots, e_n) = \operatorname{ad}^*_{e} \eta^{(k)}(e_{k+1}, \dots, e_n)\nonumber \\
 \qquad{} - \sum_{i=k}^n (-1)^{i-1} \eta^{(k)}([e, e_i], e_{k+1}, \dots, \check{e}_i, \dots, e_n),\qquad k=1, \dots, n-1, \nonumber\\ %\label{conditionofgsm34}\\
\operatorname{ad}^*_{e} \eta^{(0)}(e_1, \dots, e_n)= \sum_{i=1}^n (-1)^{i-1} \eta^{(0)}([e, e_i], e_1, \dots, \check{e}_i, \dots, e_n), \qquad k=0.\nonumber
%\label{conditionofgsm35}
\end{gather}
A pre-$n$-plectically anchored condition equation \eqref{conditionofgsm26} is trivially satisfied from equation \eqref{conditionofgsm33},
\begin{gather*}
\md \iota_{\rho} \widetilde{h} =0.%\label{conditionofgsm36}
\end{gather*}
This condition already appeared in~\cite{Kotov:2016lpx}.

The above conditions are a direct generalization of a momentum map (multimomentum map) on a multisymplectic manifold with a Lie group action \cite{Carinena:1992rb, Gotay:1997eg}
by setting $\eta^{(k)} =0$ for $k=0, \dots, n-2$. In this case, $\eta^{(n-1)}$ is a multimomentum map.

\section{Discussion and outlook}\label{section6}

We have shown that a simple constrained Hamiltonian mechanics and a two-dimensional gauged sigma model with boundary have a momentum section and a Hamiltonian Lie algebroid structure. By generalizing a gauged sigma model to a higher-dimensional gauged sigma model with WZ term, we have proposed a theory of a multimomentum section on a pre-multisymplectic manifold.

It is important to compare other generalizations of a moment map theory to a multisymplectic manifold such as Madsen--Swann's multimoment map on the $n$-th Lie kernel \cite{Madsen:2010qp, Madsen:2011ru}, a homotopy moment map \cite{Fregier:2013dda}, and a weak moment map~\cite{Herman:2018box}.

Though we proposed a momentum section on a pre-multisymplectic manifold~\eqref{conditionofgsm21} and~\eqref{conditionofgsm25} from consistency conditions of a higher-dimensional gauged nonlinear sigma model, their geometrical structures should be analyzed more. These structure are described by a Lie algebroid differential ${}^E d$ and a covariant derivative $D$.

In all examples in our paper, the pre-symplectically anchored condition (H1) is not necessary for consistency of structures. Conditions (H2) and (H3) are essential for physical applications. More examples are needed for deeper understanding of a momentum section theory.

We have assumed an anchor map $\rho$ is injective in this paper. However we should relax this condition. If an anchor map $\rho$ is not necessarily injective,
we can consider more general algebroid such as a Courant algebroid~\cite{LWX}, a Lie 3-algebroid~\cite{Ikeda:2010vz}, and higher algebroids, as a symmetry of a gauged sigma model. This direction is related to a Lie group action on a Courant algebroid and the reduction~\cite{Bursztyn:2005vwa}. These generalizations are left for future analysis.

We considered an infinitesimal version, i.e., an action of a Lie algebroid on a pre-(multi) symplectic manifold. A globalization to a Lie groupoid corresponding to a generalization of a~Lie group action is a next problem.

A next step of physical systems in this paper is quantization. One possible quantization is an equivariant localization using the Duistermaat--Heckman formula, which has been already discussed in~\cite{Blohmann:2018}. In this paper, we have obtained more concrete physical models for applications of the localization.
For this purpose, the condition (H1) looks like essential since we need an equivariant differential such that $D^2=0$.

Since a momentum section and a Hamiltonian Lie algebroid structure is a natural structure on a gauged sigma model, we can hope to obtain new physical results from analysis of a Hamiltonian Lie algebroid.

\subsection*{Acknowledgments}
The author would like to thank Yuji Hirota, Kohei Miura, Satoshi Watamura and Alan Weinstein for useful comments. He thanks the referees for their careful reading of the manuscript and especially for their helpful comments.

\pdfbookmark[1]{References}{ref}
\LastPageEnding

\end{document}